\documentclass{appolb}
\usepackage{graphicx}
\usepackage{amssymb}
\usepackage{amsmath}
\usepackage{color}
\begin{document}

\title{Saturation and Geometrical Scaling:  \\ from Deep Inelastic ep Scattering \\
to Heavy Ion Collisions
\thanks{Presented at the Conference {\em Excited QCD}, Taranska Lomnica, Slovakia, March 8 -- 14, 2015.}}
\author{Michal Praszalowicz
%\email{michal@if.uj.edu.pl}
\address{M. Smoluchowski Institute of Physics, Jagiellonian University, \\
ul. S. {\L}ojasiewicza 11, 30-348 Krak{\'o}w, Poland.}}

\maketitle

\begin{abstract}
Saturation of gluon distribution is a consequence of the
non-linear evolution equations of QCD. Saturation implies
the existence of so called saturation momentum which is defined as a
gluon density per unit rapidity per transverse area. At
large energies for certain kinematical domains saturation
momentum is the only scale for physical processes. As
a consequence different observables exhibit
geometrical scaling (GS). We  discuss a number of
examples of GS  in different reactions.
\end{abstract}

\PACS{13.85.Ni,12.38.Lg}

\section{Introduction}

\label{intro}

At the  eQCD meeting in 2013 \cite{Praszalowicz:2013swa} we have discussed the
emergence of geometrical scaling \cite{Stasto:2000er} for $F_{2}/Q^{2}$ in
deep inelastic scattering (DIS) \cite{Praszalowicz:2012zh} and for charged
particle distributions in proton collisions \cite{McLerran:2010ex}. Here,
after short reminder, we extend this analysis to $\left\langle p_{\mathrm{{T}%
}}\right\rangle (N_{\mathrm{ch}})$ correlation
\cite{McLerran:2013oju,Praszalowicz:2014kaa} and to heavy ion collisions (HI)
\cite{Praszalowicz:2011rm}. References \cite{Praszalowicz:2013swa},
\cite{Praszalowicz:2012zh}--\cite{Praszalowicz:2011rm} include a more complete
bibliography of the subject.

Geometrical scaling hypothesis means that some observable $\sigma$ that in
principle depends on two independent kinematical variables, say $x$ and
$Q^{2}$, in fact depends only on a specific combination of them denoted as
$\tau$:
\begin{equation}
\sigma(x,Q^{2})=S_{\bot} F(\tau). \label{GSdef}%
\end{equation}
Here function $F$ in Eq.~(\ref{GSdef}) is a dimensionless function of scaling
variable
\begin{equation}
\tau=Q^{2}/Q_{\text{s}}^{2}(x). \label{taudef}%
\end{equation}
and
\begin{equation}
Q_{\text{s}}^{2}(x)=Q_{0}^{2}\left(  {x}/{x_{0}}\right)  ^{-\lambda}
\label{Qsat}%
\end{equation}
is the saturation scale. $S_{\bot}$ is a transverse area that corresponds to
the overlap of hadrons colliding at fixed impact parameter $b$ (or integrated
over $db$), or -- like in the case of DIS -- it is a cross section for large
dipole scattering on a proton. $Q_{0}$ and $x_{0}$ in Eq.~(\ref{Qsat}) are
free parameters, which can be extracted from the data within some specific
model for $\sigma$, and parameter $\lambda$ is a dynamical quantity of the
order of $\lambda\sim0.3$. Here we shall test the hypothesis whether different
pieces of data can be described by formula (\ref{GSdef}) with \emph{constant}
$\lambda$, and what is the range of transverse momenta where GS is working
satisfactorily. Throughout this paper we shall be neglecting logarithmic
energy dependence due to the running of $\alpha_{\mathrm{s}}$.

\section{Deep Inelastic Scattering at HERA}

\label{DIS}

\begin{figure}[h]
\centering
\includegraphics[width=6.24cm,angle=0]{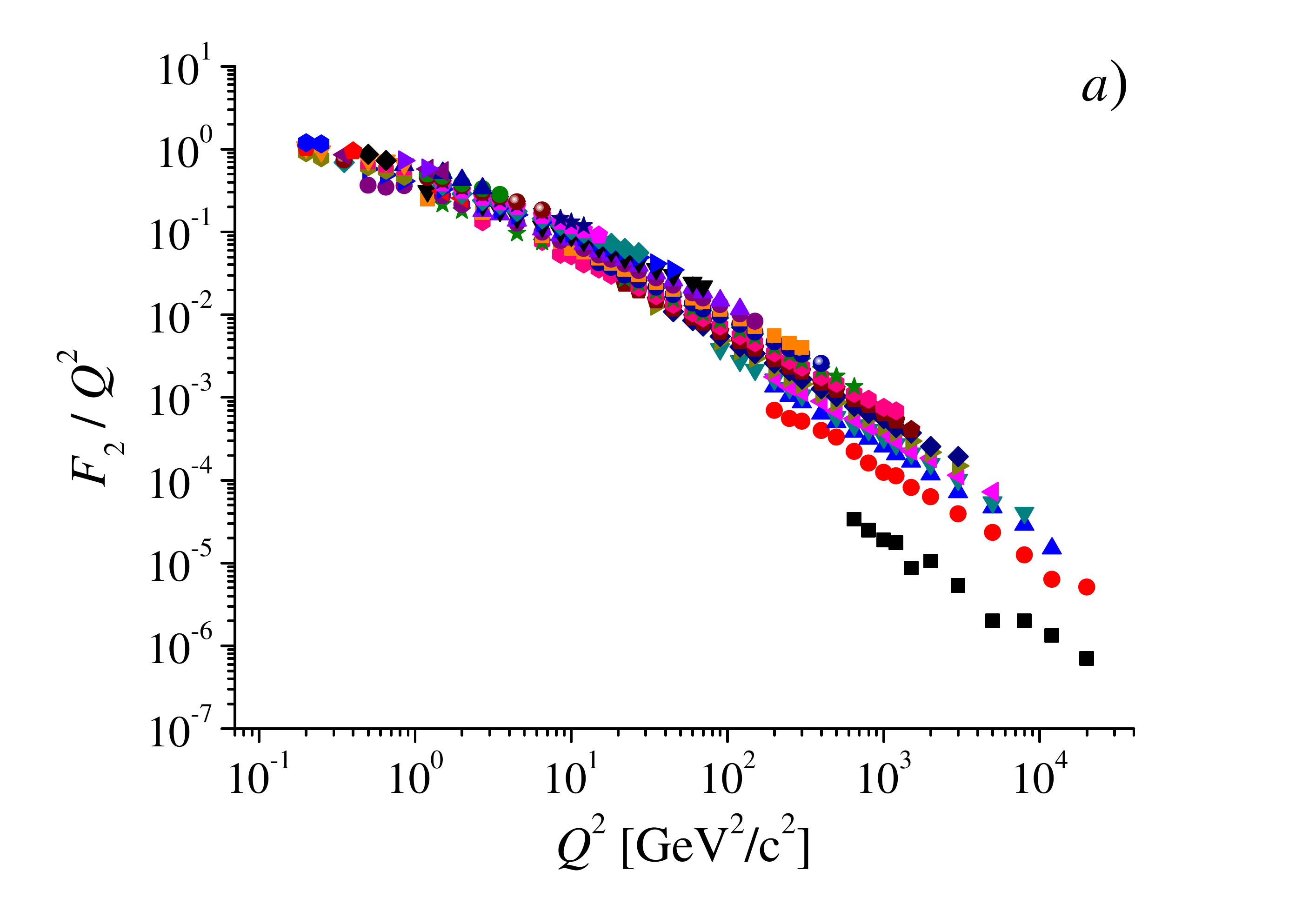}
\includegraphics[width=6.24cm,angle=0]{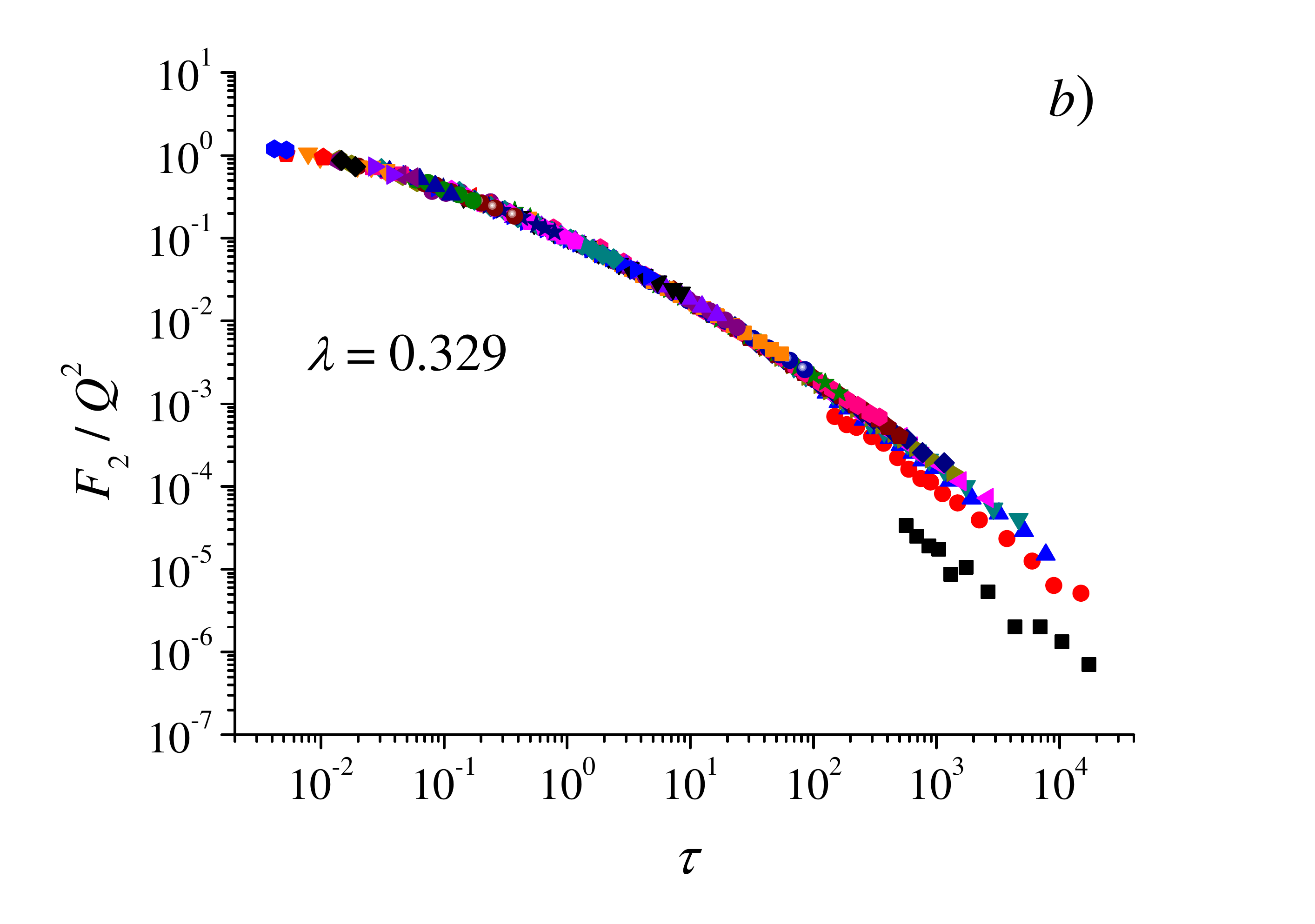} \caption{Combined DIS data
\cite{HERAdata} for $F_{2}/Q^{2}$. Different points forming a wide band as a
function of $Q^{2}$ in the left panel correspond to different Bjorken $x$'s.
They fall on a universal curve when plotted in terms of $\tau$ (right panel).
(Figure from the first paper of Ref.~\cite{Praszalowicz:2012zh}).}%
\label{GSDIS}%
\end{figure}

Let us start with DIS where the relevant scaling observable is $F_{2}%
(x)/Q^{2}$~\cite{Stasto:2000er}. In Fig.~\ref{GSDIS} we plot $F_{2}(x)/Q^{2}$
as a function of $Q^{2}$ (left panel) and in terms of $\tau$ for
$\lambda=0.329$ (right panel) for combined HERA data \cite{HERAdata}.
Different points correspond to different Bjorken $x$'s. We see from
Fig.~\ref{GSDIS} that points of different Bjorken $x$'s scale very well with
some exception in the right part of Fig.~\ref{GSDIS}.b. These points, however,
correspond to large Bjorken $x$'s where GS is supposed to break.

\section{Inelastic $p_{\mathrm{T}}$ spectra at the LHC}

\label{ppLHC}

In hadronic collisions at c.m. energy $W=\sqrt{s}$ particles are produced in
the scattering process of two patrons characterized by Bjorken $x$'s
\begin{equation}
x_{1,2}=e^{\pm y}\,p_{\text{T}}/W. \label{x12}%
\end{equation}
For central rapidities $x=x_{1} \sim x_{2}$. Geometrical scaling in this case
means simply that~\cite{McLerran:2010ex}:
\begin{equation}
\left.  \frac{dN}{dy d^{2}p_{\text{T}}}\right\vert _{y\simeq0}= S_{\bot}
F(\tau) \label{GSinpp}%
\end{equation}
where $F$ is a universal dimensionless function of the scaling variable
\begin{equation}
\tau=p_{\text{T}}^{2}/Q_{\text{s}}^{2}(x)= p_{\text{T}}^{2}/Q_{0}^{2}
\,\left(  p_{\mathrm{T}}/(x_{0} W)\right) ^{\lambda}. \label{taudef1}%
\end{equation}

\begin{figure}[h]
\centering
\includegraphics[width=6cm,angle=0]{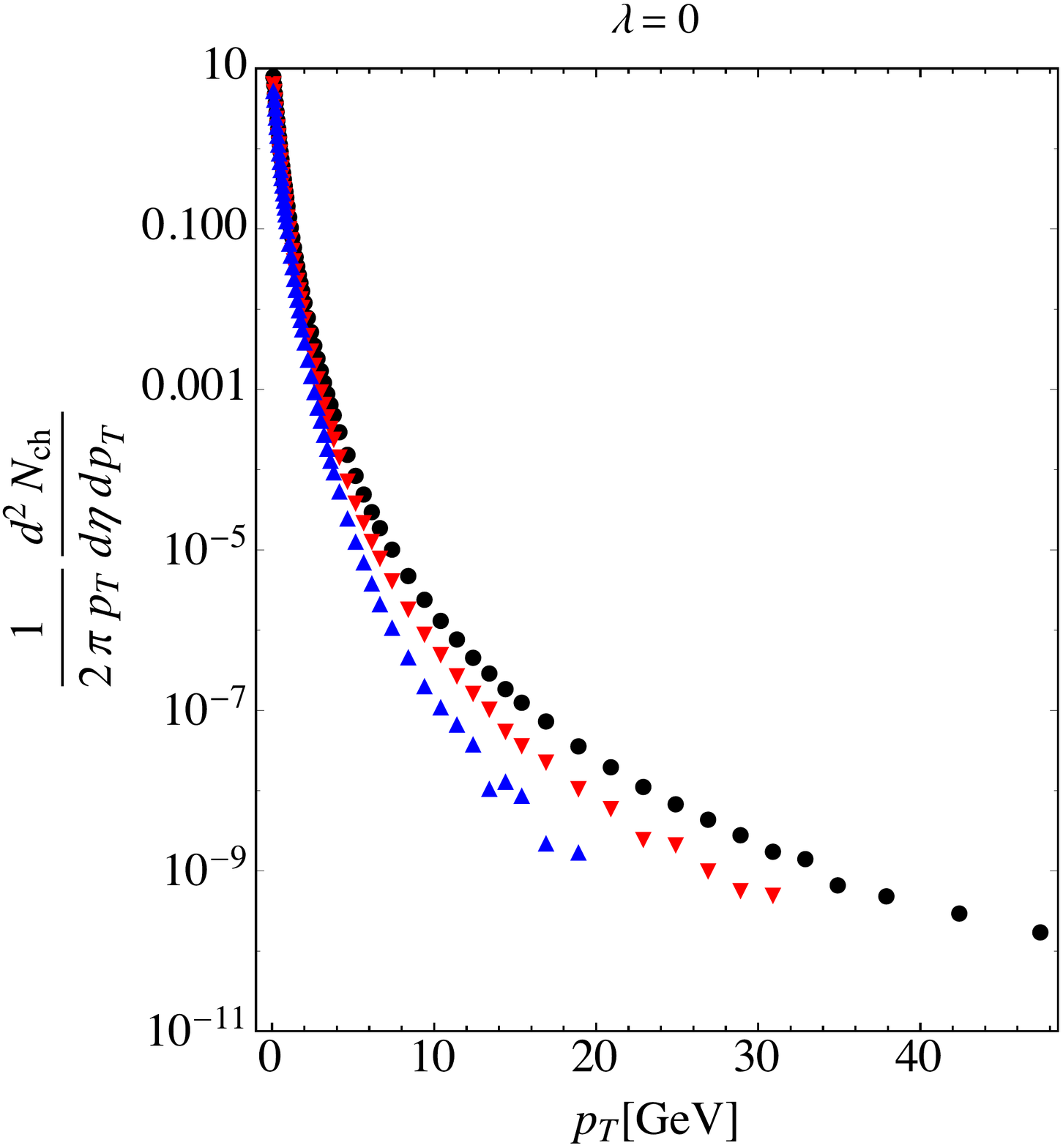}
\includegraphics[width=6cm,angle=0]{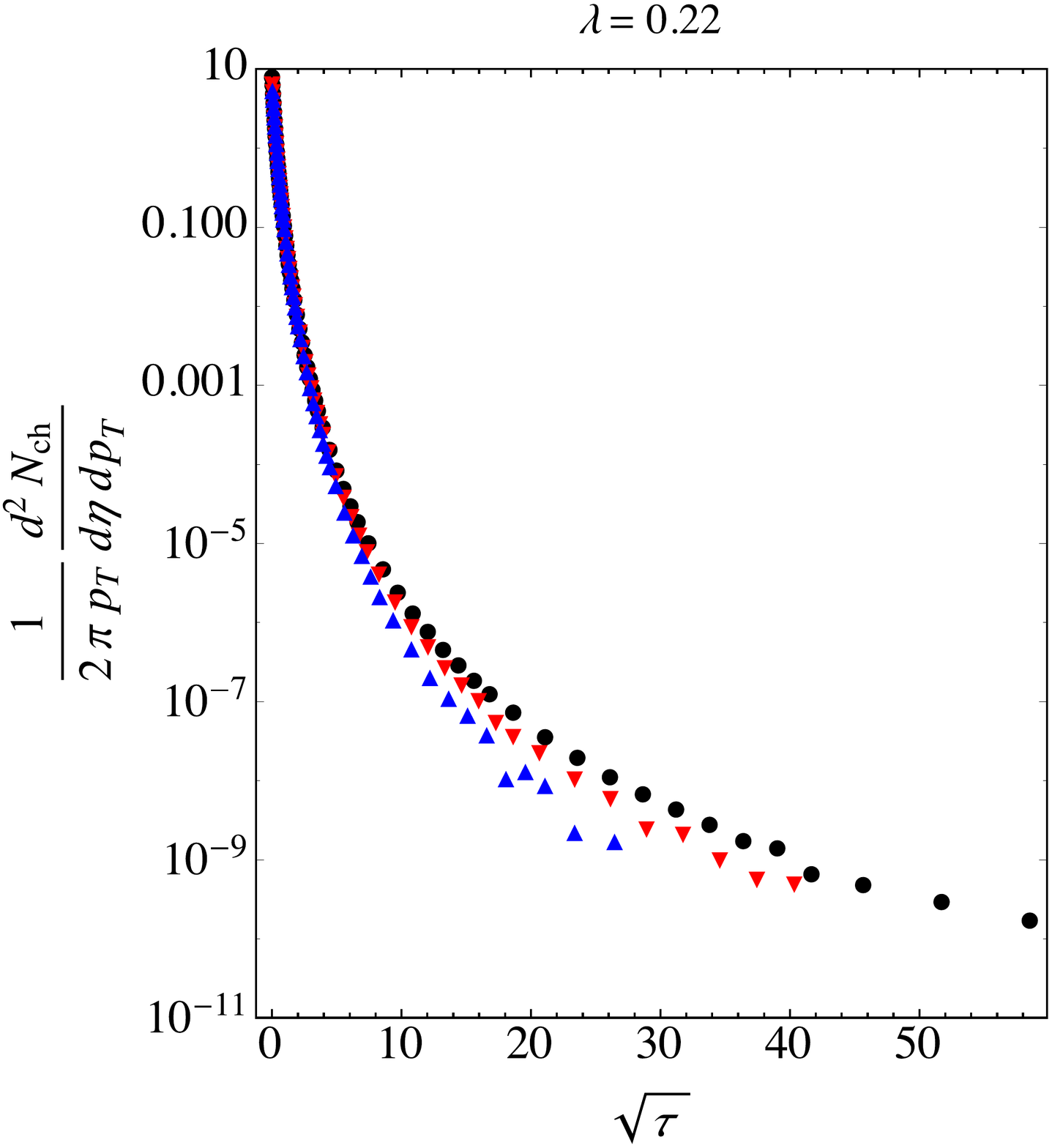} \caption{Data
for pp scattering from ALICE \cite{Abelev:2013ala} plotted in terms of
$p_{\mathrm{T}}$ and $\sqrt{\tau}$. Full (black) circles correspond to
$W=7$~TeV, down (red) triangles to 2.76~TeV and up (blue) triangles to
0.9~TeV.}%
\label{GSALICE}%
\end{figure}

In Fig.~\ref{GSALICE} we plot ALICE pp data \cite{Abelev:2013ala} in terms of
$p_{\mathrm{T}}$ (left panel) and in terms of scaling variable $\tau$ (right
panel) for $\lambda=0.22$. We have found by a model independent analysis that
the optimal exponent $\lambda=0.22-0.24$ \cite{FraPra}, which is smaller than
in the case of DIS. Why this so, remains to be understood.

An immediate consequence of GS for the $p_{\mathrm{T}}$ spectra is a
power-like growth of multiplicity with energy. Indeed, since
\begin{equation}
p_{\text{T}}=\overline{Q}_{\text{s}}(W)\tau^{1/(2+\lambda)}\label{pT}%
\end{equation}
where the \emph{average} saturation scale is defined as%
\begin{equation}
\overline{Q}_{\text{s}}(W)=Q_{0}\left(  {x_{0} W}/{Q_{0}}\right)
^{\lambda/(2+\lambda)}\label{Qbarsat}%
\end{equation}
one arrives at
\begin{equation}
\frac{dN}{dy}=S_{\bot} \overline{Q}_{\text{s}}^{2}(W) \times\left[  \frac
{1}{2+\lambda}%
%TCIMACRO{\dint }%
%BeginExpansion
{\displaystyle\int}
%EndExpansion
\mathcal{F}(\tau)\tau^{2/(2+\lambda)}\frac{d\tau}{\tau} \right] .
\end{equation}
Data indeed support the power-like growth of inelastic multiplicity as
$s^{0.1}$ as predicted by GS by Eq.~(\ref{Qbarsat}) for $\lambda=0.22-0.24$.

\section{Mean $p_{\mathrm{T}}$ in hadronic collisions at the LHC}

\label{meanpT}

\begin{figure}[h]
\centering
\includegraphics[height=.25\textheight]{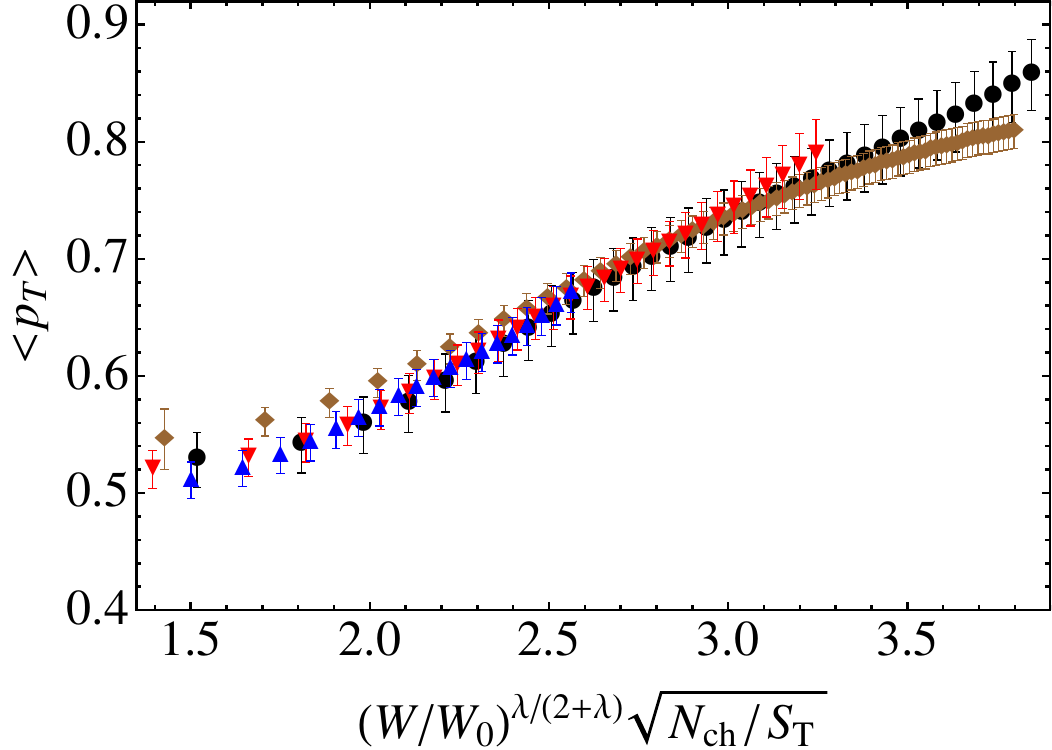}\caption{Mean
$\left\langle p_{\mathrm{{T}}}\right\rangle $ in pp collisions at 7 TeV (full
black circles), 2.76 TeV (full red down-triangles), 0.9 TeV (full blue
up-triangles) and in pPb collisions at 5.02 TeV (full brown diamonds) plotted
in terms of scaling variable $(W/W_{0})^{\lambda/(2+\lambda)}\sqrt
{N_{\mathrm{ch}}/ S_{\bot}}$. For pp $W_{0}=7$~TeV and for pPb $W_{0}%
=5.02$~TeV. (Figure from the second paper of Ref.~\cite{McLerran:2013oju}.)}%
\label{fig:scaled}%
\end{figure}

Another consequence of Eq.~(\ref{GSinpp}) is that \cite{McLerran:2013oju}
\begin{equation}
\left\langle p_{\mathrm{{T}}}\right\rangle \sim\bar{Q}_{\mathrm{s}}(W) \,
,\label{meanpT1}%
\end{equation}
which means that $\left\langle p_{\mathrm{{T}}}\right\rangle $ rises with
energy as $W^{\lambda/(2+\lambda)}$, which is indeed seen in the data. On the
other hand, since the saturation momentum is by Eq.~(\ref{Qbarsat}) equal to
the gluon density per transverse area, one easily derive the correlation
between mean $p_{\mathrm{T}}$ and charged particles multiplicity at given
energy $W$ \cite{McLerran:2013oju}:
\begin{equation}
\left.  \left\langle p_{\text{T}}\right\rangle \right\vert _{W} \sim\left(
\frac{W}{W_{0}}\right)  ^{\lambda/(2+\lambda)}\sqrt{\frac{N_{\text{ch}}%
}{\left.  S_{\bot}(N_{\text{ch}})\right\vert _{W_{0}}}}\, . \label{pTWdep}%
\end{equation}
By fixing multiplicity, one is probing some fixed impact parameter that
corresponds to the overlap transverse area $S_{\bot}(N_{\text{ch}})$ that
itself is by construction both multiplicity and energy dependent. Therefore
one needs a model for $S_{\bot}(N_{\text{ch}})$. To this end we have used the
Color Glass Condensate result for pp and pA collisions \cite{Bzdak:2013zma}.
The result is plotted in Fig.~\ref{fig:scaled} where we plot ALICE data
\cite{Abelev:2013bla} as a function of scaling variable defined in
Eq.~(\ref{pTWdep})

\section{Geometrical Scaling in heavy ion collisions}

\label{sec:HI}

\begin{figure}[h]
\centering
\includegraphics[width=6.1cm]{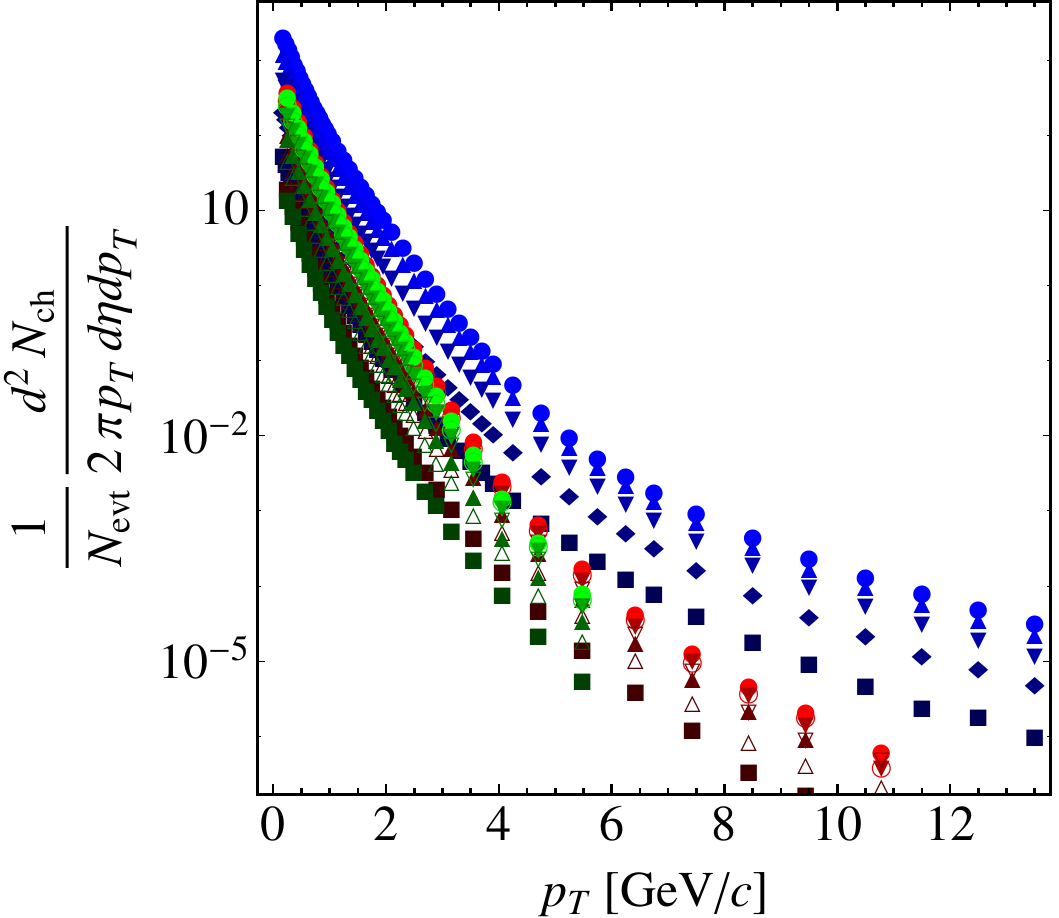}
\includegraphics[width=6.1cm]{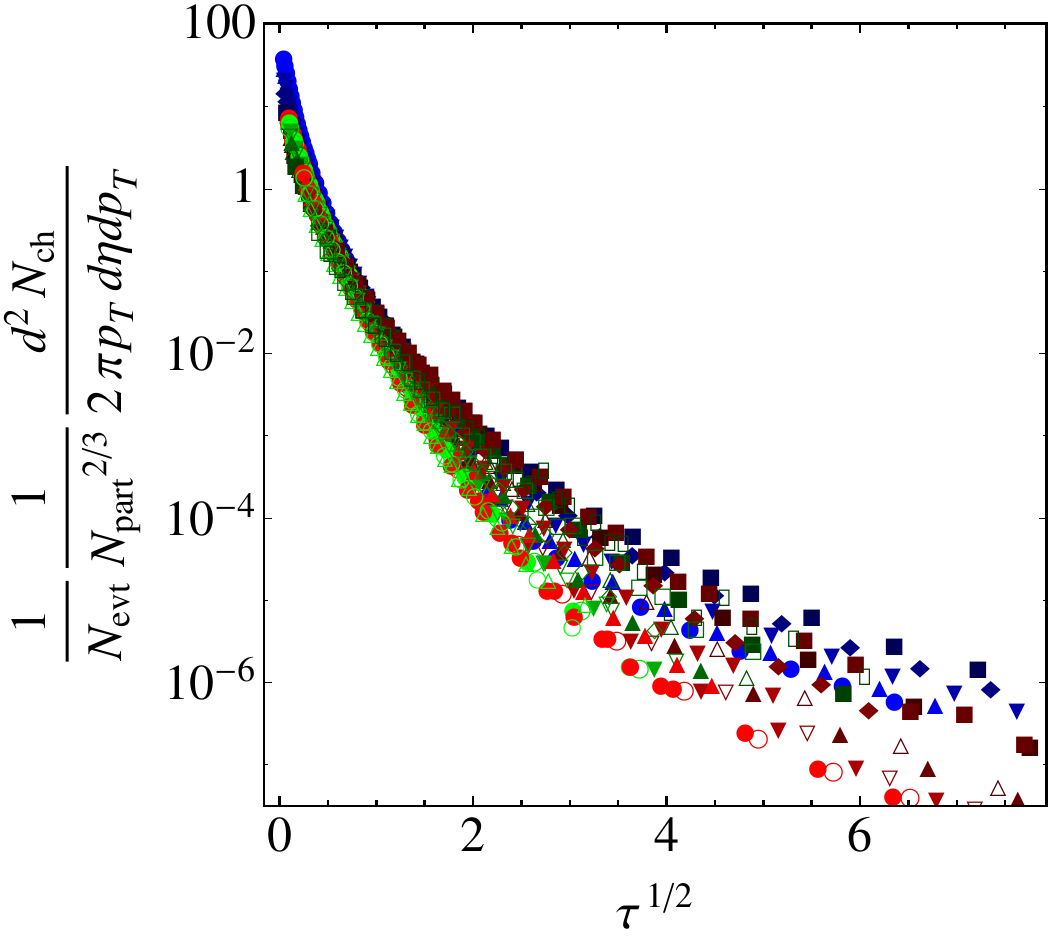}\caption{Illustration of
geometrical scaling in heavy ion collisions at different energies and
different centrality classes. Left panel shows charged particle distributions
from ALICE \cite{Abelev:2012hxa}, STAR \cite{Adams:2003kv,Adler:2002xw} and
PHENIX \cite{Adler:2003au,Adcox:2001jp} plotted as functions of $p_{\mathrm{T}%
}$. In the right panel the same distributions are scaled according to
Eq.~(\ref{multHI}). }%
\label{fig:all}%
\end{figure}

GS for particle spectra in HI collisions has been already discussed in
Ref.~\cite{Praszalowicz:2011rm} and in Ref.~\cite{Klein-Bosing:2014uaa} for
photons. HI data are divided into centrality classes that select events within
certain range of impact parameter $b$. In this case both transverse area
$S_{\bot}$ and the saturation scale $Q_{\text{s}}^{2}$ acquire additional
dependence on centrality that is characterized by an average number of
participants $N_{\text{part}}$. We have
\cite{Klein-Bosing:2014uaa,Kharzeev:2004if}:%
\begin{equation}
S_{\bot}\sim N_{\text{part}}^{2/3}\;\text{and}\;Q_{\text{s}}^{2}\sim
N_{\text{part}}^{1/3}. \label{Npartscaling}%
\end{equation}
Therefore in HI collisions%
\begin{equation}
\frac{1}{N_{\text{evt}}}\frac{dN_{\text{ch}}}{N_{\text{part}}^{2/3} \,d\eta
d^{2}p_{\text{T}}}=\, \frac{1}{Q_{0}^{2}} \, F(\tau) \,\,\,\, \mathrm{where}
\,\,\,\, \tau=\frac{p_{\text{T}}^{2}}{N_{\text{part}}^{1/3} \, Q_{0}^{2}%
}\left(  \frac{p_{\text{T}}}{W}\right)  ^{\lambda}. \label{multHI}%
\end{equation}

In Fig.~\ref{fig:all} we plot LHC and RHIC data in terms of $p_{\rm T}$ (left
panel) and $\sqrt{\tau}$ for $\lambda=0.3$ (right panel). One can see an
approximate scaling of, however, worse quality than in the pp case.

\bigskip

To summarize: a wealth of data in hadronic collisions exhibit GS. This may be
interpreted as a signature of saturation. However some details, like the
non-universality of the value of $\lambda$, remain to be understood.

%\section*{Acknowledgemens}
\bigskip\noindent This work was supported by the Polish NCN grant 2014/13/B/ST2/02486.

\end{document}